\documentclass[twocolumn,superscriptaddress,aps,prb]{revtex4-1}
\usepackage[latin9]{inputenc}
\usepackage{color}
\usepackage{graphicx}
\usepackage{esint}



\begin{document}

\title{Optomagnonic whispering gallery microresonators}

\author{Xufeng Zhang}
\affiliation{Department of Electrical
Engineering, Yale University, New Haven,
Connecticut 06511, USA}

\author{Na Zhu}
\affiliation{Department of Electrical
Engineering, Yale University, New Haven,
Connecticut 06511, USA}

\author{Chang-Ling Zou}
\affiliation{Department of Electrical
Engineering, Yale University, New Haven,
Connecticut 06511, USA}

\author{Hong X. Tang}
\email{corresponding email: hong.tang@yale.edu}
\affiliation{Department of Electrical
Engineering, Yale University, New Haven,
Connecticut 06511, USA}

\date{\today}

\begin{abstract}
Magnons in ferrimagnetic insulators such as
yttrium iron garnet (YIG) have recently emerged
as promising candidates for coherent information
processing in microwave circuits. Here we
demonstrate optical whispering gallery modes of a
YIG sphere interrogated by a silicon nitride
photonic waveguide, with quality factors
approaching $10^{6}$ in the telecom c-band after
surface treatments. Moreover, in contrast to
conventional Faraday setup, this implementation
allows input photon polarized colinearly to the
magnetization to be scattered to a sideband mode
of orthogonal polarization. This Brillouin
scattering process is enhanced through triply
resonant magnon, pump and signal photon modes -
all of whispering gallery nature - within an
``optomagnonic cavity''. Our results show the
potential use of magnons for mediating
microwave-to-optical carrier conversion.
\end{abstract}

\maketitle

Hybrid magnonic systems have been emerging
recently as an important approach towards
coherent information processing
\cite{Chumak2015,Tabuchi2015, CanMing2015,
Tabuchi2014_PRL, Zhang2014_PRL, Goryachev2014,
Bai2015, Huebl2013, Zhang2015}. The building
block of such systems, magnon, is the quantized
magnetization excitation in magnetic materials
\cite{Serga2010,Lenk2011}. Its great tunability
and long lifetime make magnon an ideal
information carrier. Particularly, in magnetic
insulator yttrium iron garnet (YIG), magnons
interact with microwave photons through magnetic
dipole interaction, which can reach the strong
and even ultrastrong coupling regime thanks to
the large spin density in YIG
\cite{Tabuchi2014_PRL,Zhang2014_PRL,Goryachev2014}.
Besides, the magnon can also couple with the
elastic wave \cite{Kittel1958,Sinha1962} and
optical light \cite{Shen1966,Demokritov2001}, it
is of great potential as an information
transducer that mediates inter-conversion among
microwave photon, optical photon and acoustic
phonon. Long desired functions, such as
microwave-to-optical conversion, can be realized
on such a versatile platform.

Magneto-optical (MO) effects such as Faraday
effect have been long discovered and utilized in
discrete optical device applications
\cite{Freiser1968,Sugano,Bi2011}. Based on such
effects, magnons can coherently interact with
optical photons. On the one hand, magnon can be
generated by optical pumps
\cite{Kirilyuk2010,Kimel2005,Satoh2012,Ziel1965}.
On the other hand, optical photons can be used to
probe magnon through Brillouin light scattering
(BLS) \cite{Fleury1968,Demokritov2001}. However,
in previous studies the typical geometries are
all thin film or bulk samples inside which the
optical photon interacts with magnon very weakly,
usually only through a single pass. For high
efficient magnon-photon interaction, it is
desirable to obtain triple resonance condition of
high quality ($Q$) factor modes, i.e., the
magnon, the input and the output optical photons
are simultaneously on resonance.

In this Letter, we demonstrate the magnon-photon
interaction in a high $Q$ optomagnonic cavity
which simultaneously supports whispering gallery
modes (WGMs) of optical and magnon resonances.
With high-precision fabrication and careful
surface treatment, the widely used YIG sphere
structure, which is inherently an excellent
magnonic resonator, exhibits high optical $Q$
factors in our measurements. YIG has a high
refractive index (2.2 in the telecom c-band),
which poses a challenge for efficient light
coupling with silica fiber tapers. By employing
an integrated silicon nitride optical waveguide
with an effective index (around 2.0) matching
that of YIG, we can efficiently couple to both
the TM and TE optical resonances in the YIG
sphere. By driving the system with an optical
pump, the magnons excited by an input microwave
signal can be converted into optical signal of a
different color. Our demonstration shows the
great potential of YIG sphere as a platform to
bridge the gap between magnon and optical
photons, paving the way towards using magnon as a
transducer for coherent information processing
between distinct carriers.

In magnetized YIG material, magnons are the
collective excitations of spin states of
Fe$^{3+}$ ions \cite{Stancil2009}. The creation
or annihilation of a single magnon corresponds to
the ground spin flip. At the microwave
frequencies, the magnon state can be manipulated
straightforwardly by magnetic dipolar transition
using the oscillating magnetic fields of
microwave photons. While at the optical
frequencies, the magnon manipulation becomes
difficult because the magnetic transition is
negligible whilst direct spin-flip by electric
dipole transition is forbidden \cite{LEGALL1971}.
Alternatively, the optical photons can modify the
ground state spin through a two-photon transition
by means of an orbital transition and spin-orbit
interaction (the MO effect) \cite{Shen1966}. Such
a process has been previously studied using
conventional Faraday setups
\cite{Borovik-Romanov1982}, in which light
propagates parallel to the magnetic field and
interacts with magnon in a single pass. It is
natural to consider shaping the YIG into an
optical cavity to boost up the magnon-photon
interaction as light passes the magnetic material
multiple times. Therefore, we propose to use a
whispering gallery resonator made by YIG to
provide enhanced magnon-photon coupling in a
triply resonant configuration, and the mechanism
is explained in the following discussions.

\begin{figure}[htpb]
\begin{centering}
\includegraphics[width=0.475\textwidth]{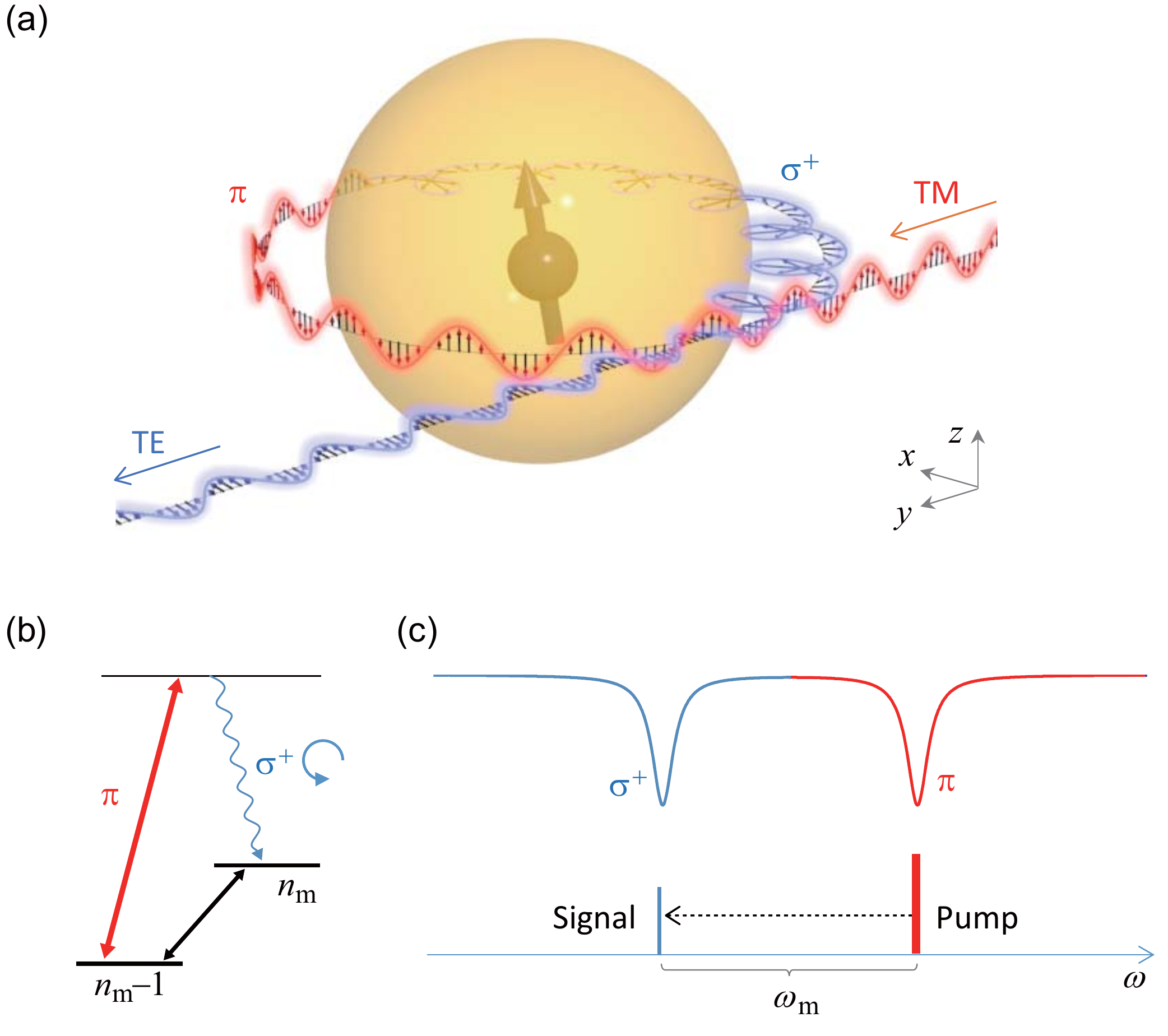}
\protect\caption{(a) Schematic illustration of
the magnon-photon interaction. The YIG sphere is
biased by a magnetic field along $z$ direction,
while the WGMs propagate along the perimeter in
the $x$-$y$ plane. The TM input light excites the
$\pi$ WGM in the YIG sphere, which is scattered
by magnon into $\sigma^{+}$ polarized photon and
then converts to the TE output in the waveguide.
(b) Energy level diagram of the magnon-photon
interaction. (c) Triple resonance condition for
the enhanced magnon-photon interaction process in
the optomagnonic resonator. }
\label{fig:mechanism}
\par\end{centering}

\end{figure}

The optomagnonic cavity we used in our
experiments is a single crystal YIG sphere. Due
to the spherical symmetry, lights are confined in
the sphere by total internal reflection and form
WGMs. Each optical WGM is characterized by three
mode numbers $(q,l,n)$, which correspond to the
radial, angular and azimuthal order
($n=-l,\ldots,l$), respectively
\cite{Braginsky1989, Chiasera2010}. Moreover, the
WGMs are also characterized by their
polarization, i.e., the direction of their
electric field distribution. Conventional Faraday
setups require the bias magnetic field to be
parallel to the direction of light propagation.
However, for WGMs light propagates along the
circumference of the sphere
{[}Fig.\,\ref{fig:mechanism}(a){]}, therefore the
bias magnetic field should be in $x$-$y$ plane.
Due to the geometry symmetry, the MO effect
vanishes for such a Faraday configuration. At a
first glance, the MO effect also vanishes for
bias magnetic field along $z$ direction, since it
requires circular polarization in respect to
$\overrightarrow{H}$ while WGMs are linearly
polarized, either parallel (TE) or perpendicular
(TM) to the $z$ direction. However, thanks to the
field gradient at the dielectric interface
\cite{Junge2013,Petersen2014,Shomroni2014}, there
are non-zero optical electric fields along the
propagation direction for the TE polarized WGMs.
As a result, the electric field rotates within
the $x$-$y$ plane and forms a cycloid trajectory,
similar to the elliptically polarized light
propagating in free space. Therefore, the TE WGMs
possess partial circular polarization
($\sigma^{+}$) and can have magnetic response via
Faraday effect, as schematically illustrated in
Fig.\,\ref{fig:mechanism}(a). Note that the pump
light can propagate either clockwise (CW) or
counterclockwise (CCW), with different
conservation conditions accordingly, as will be
shown below.

\begin{figure}[t]
\begin{centering}
\includegraphics[width=0.475\textwidth]{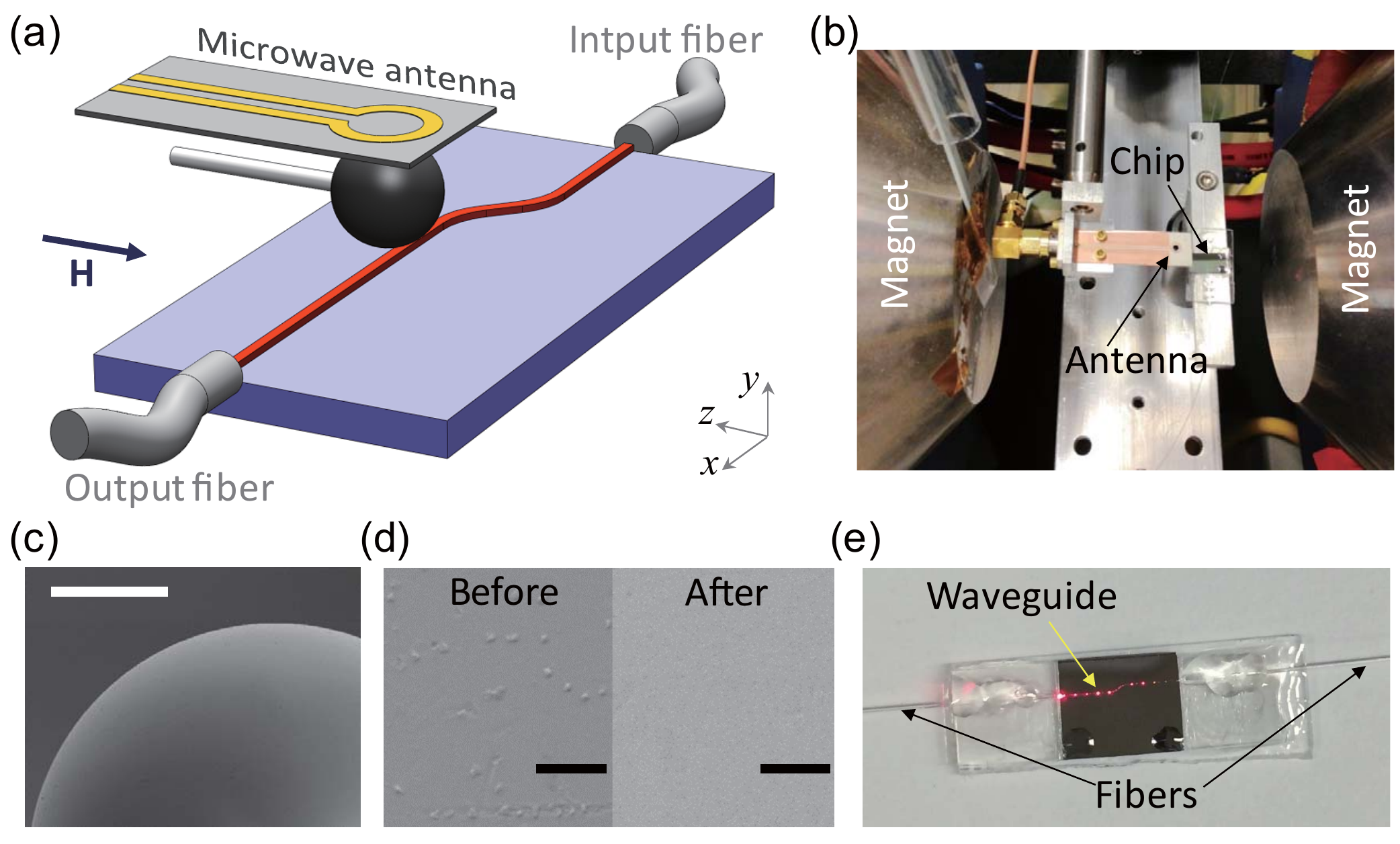}
\protect\caption{(a) and (b) Schematic and
optical image of the experimental assembly
of our optomagnonic device, respectively. (c)
Scanning electron microscope image of the
polished YIG sphere. The scale bar is 100 $\mu$m.
(d) The surface of the YIG sphere before and
after our surface treatment process. Scale bars
are 1 $\mu$m. The sub-micrometer particles vanish
after the surface treatment. (e) Optical image of
the silicon nitride coupling waveguide chip with
glued fibers on the two sides. The chip and the
fibers are attached to a piece of glass holder
for mechanical support and reducing long-term
drift.}\label{fig:device}

\par\end{centering}

\end{figure}

Similar to the optical WGMs, the magnon modes in
YIG sphere can also be characterized by three
mode numbers
$(q_{\mathrm{m}},l_{\mathrm{m}},n_{\mathrm{m}})$
\cite{Roschmann1977}. For the uniform magnon mode
with all the spins precessing in phase, the
corresponding mode numbers are $(1,1,1)$. The
microscopic mechanism of the magnon-photon
interaction is intrinsically a three-wave
process, as schematically illustrated by
Fig.\,\ref{fig:mechanism}(b). Due to the spin
angular momentum conservation, every time when
the magnon number increases by $1$ it indicates
that the electron spin increases by $1$, which
corresponds to a two-photon transition in the
form of $\sigma^{+}\rightarrow\pi$ (CCW) or
$\pi\rightarrow\sigma^{-}$ (CW). As a result,
there would be only one optical sideband
generated for a given pumping light direction.
The mesoscopic model of the MO effect is
represented by the permittivity tensor
$\varepsilon_{ij}=\varepsilon_{0}(\varepsilon_{r}\delta_{ij}-if\epsilon_{ijk}M_{k})$
\cite{Stancil2009}, where $\varepsilon_{0}$ is
the vacuum permittivity, $\varepsilon_{r}$ is the
relative permittivity of YIG, $\delta_{ij}$ and
$\epsilon_{ijk}$ are Kronecker and Levi-Civita
symbols, $f$ is the Faraday coefficient, $M_{k}$
is the magnetization, and $i$, $j$, $k$
correspond to $x$, $y$, $z$ direction,
respectively. When the energy is conserved for
the two-photon and magnon transitions that
$\omega_{1}-\omega_{2}=\omega_{\mathrm{m}}$, the
coupling strength between two optical modes is
$g=\int\Delta\varepsilon_{ij}(\overrightarrow{x})E_{1,i}^{*}(\overrightarrow{x})E_{2,j}(\overrightarrow{x})d\overrightarrow{x}{}^{3}$,
where $E_{p,i}(\overrightarrow{x})$ ($p=1,2$) is
the normalized field of optical WGM
$\int\varepsilon_{ii}(\overrightarrow{x})|E_{p,i}(\overrightarrow{x})|^{2}d\overrightarrow{x}{}^{3}=\omega_{p}$,
and magnon induced permittivity change is
$\Delta\varepsilon_{ij}(\overrightarrow{x})=-if\varepsilon_{0}\epsilon_{ijk}M_{k}(\overrightarrow{x})$.
As the field distributions are in the form of
$e^{in\phi}$ in the spherical coordinate along
the azimuthal direction, $g$ is non-zero only for
the conservation of orbit angular momentum
$n_{1}-n_{2}=n_{\mathrm{m}}$. Therefore, when the
energy, spin and orbit angular momentum
conservation relations, i.e., the triple
resonance condition
[Fig.\,\ref{fig:mechanism}(c)] and selection rule
for our optomagnonic resonator, are
simultaneously satisfied, the coupling strength
$g$ can be greatly enhanced.

The schematic and optical images of the
experiment assembly of out optomagnonic cavity
integrated with photonic and microwave circuits
are shown in Figs.\,\ref{fig:device}(a) and (b),
respectively. A 300-$\mu$m-diameter single
crystal YIG sphere
{[}Fig.\,\ref{fig:device}(c){]} is glued to a
125-$\mu$m-diameter supporting silica fiber.
Although YIG spheres have been widely used as
magnon resonators, their potential as optical
high-$Q$ WGM microresonators has been overlooked.
In fact, the low absorption loss of YIG in the
infrared wavelengths (0.13 dB/cm) \cite{Wood1967}
can lead to $Q$ factors as high as
$3\times10^{6}$. Nonetheless, the surface defects
and contamination of commercial YIG sphere
products induce strong scattering losses,
limiting the highest achievable $Q$ factor in our
experiment. A major contribution of the surface
contamination is the residual of the
sub-micrometer aluminum oxide polishing grit used
in the YIG sphere production process, which is
very difficult to remove using conventional
cleaning procedures. By combining a mechanical
polishing procedure (using silicon oxide slurry)
and a follow-up chemical cleaning procedure
(using buffered oxide etch), we efficiently
removed these contamination and obtained very
clean sphere surface
{[}Fig.\,\ref{fig:device}(d){]}. To excite the
high-$Q$ WGMs, conventional tapered silica fiber
(refractive index 1.44) approach cannot achieve
high efficiency because of the index mismatch
\cite{Zou2008}. Therefore, we integrate the YIG
sphere with a silicon nitride optical circuit
{[}Fig.\,\ref{fig:device}(e){]}, whose waveguide
mode index matches that of YIG. The chip is glued
to silica optical fibers using UV curable epoxy
after careful alignment, which provides high
efficiency and stable transmission. Another
coplanar loop antenna circuit is placed in
vicinity of the YIG sphere to convert microwave
signal to magnon. In our experiments, the YIG
sphere is always biased by an external magnetic
field along the supporting fiber ($z$) direction
according to the spin conservation condition
discussed above.

\begin{figure}[t]
\begin{centering}
\includegraphics[width=0.475\textwidth]{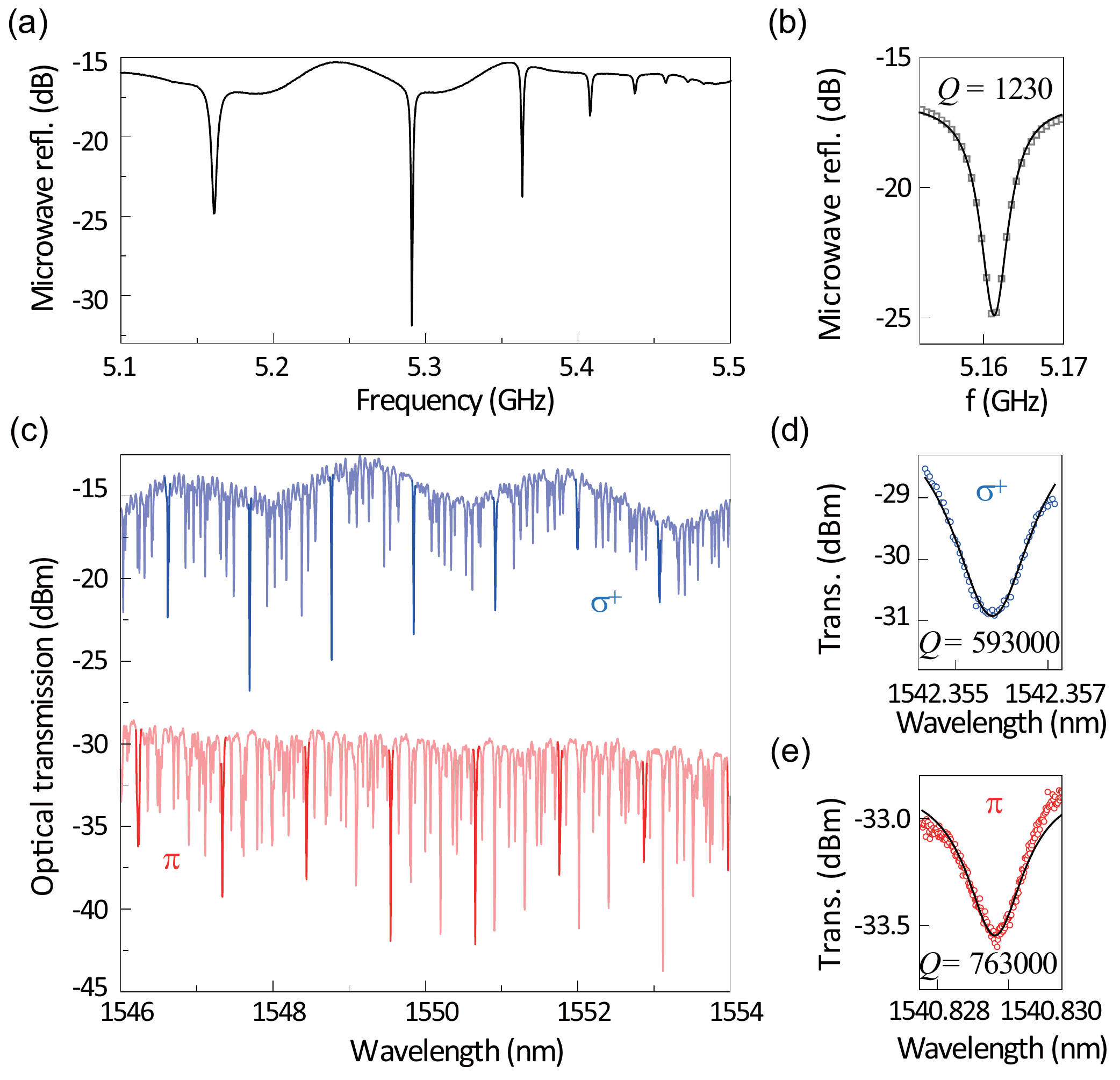}
\protect\caption{(a) Magnon resonances measured on a 300-$\mu$m-diameter YIG sphere
biased at 1840 Oe. (b) The zoomed-in spectrum of
the fundamental magnon mode. (c) Optical WGMs for
both polarizations ($\sigma^{+}$ and $\pi$)
measured on the same YIG sphere using the silicon
nitride coupling waveguide. Large extinction
ratio and the periodic mode distribution is
evident. (d) and (e) are the zoomed-in spectrum
for the two polarizations,
respectively.}\label{fig:Q}

\par\end{centering}
\end{figure}

Before studying the magnon-photon interaction, we
first characterize the optical and magnon modes.
The reflection microwave spectrum at $H=1840$ Oe
is plotted in Fig.\,\ref{fig:Q}(a), showing
multiple dips that correspond to magnon modes (to
observe high order modes, the YIG sphere is
placed at the non-uniform fields of the antenna).
In the zoomed-in spectrum of
Fig.\,\ref{fig:Q}(b), the loaded $Q$ factor of
the fundamental magnon mode $(1,1,1)$ is $1230$.
In the following magnon-photon interaction
measurement, the YIG sphere is placed at the
uniform microwave fields of the antenna output
such that only the $(1,1,1)$ mode is excited. The
optical transmission spectra are plotted in
Fig.\,\ref{fig:Q}(c), where TE/TM polarized light
in the waveguide are used to probe
$\sigma^{+}$/$\pi$ polarized WGMs in the YIG
sphere, receptively. Groups of optical resonances
show up in the spectra, exhibiting large
extinction ratio (beyond 10 dB) for both
polarizations, which confirms the efficient
coupling between the silicon nitride waveguide
and the WGMs. The measured free spectral ranges
for both the $\sigma^{+}$ (1.0765 nm) and $\pi$
(1.1068 nm) polarization agree with the
prediction (1.1580 nm) for WGMs. Thanks to our
surface treatments, very high optical $Q$ factors
are achieved: $Q_{\sigma^{+}}=0.593\times10^{6}$
and $Q_{\mathrm{\pi}}=0.763\times10^{6}$
{[}Figs.\,\ref{fig:Q}(d) and (e){]}.

\begin{figure}[t]
\begin{centering}
\includegraphics[width=0.475\textwidth]{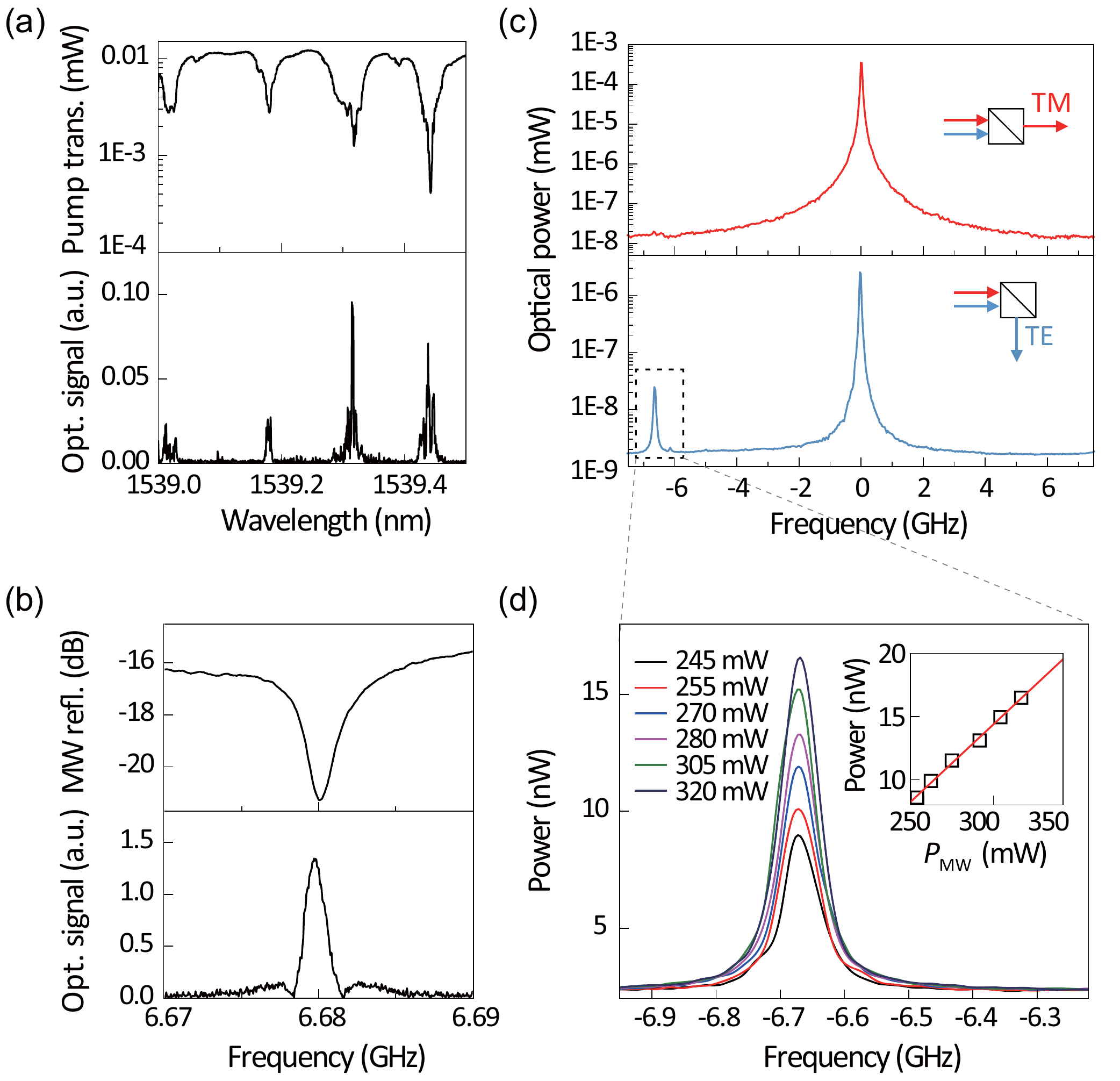}
\protect\protect\protect\caption{(a) Optical pump
transmission and the generated optical signal as
a function of pump laser wavelength. The
correspondence of the generated optical signal
peak and the optical pump resonance dip indicates
the satisfaction of the conservation conditions.
(b) Microwave reflection and the generated
optical signal as a function of the microwave
frequency. (c) Optical spectrum of the device
output when the triple resonance condition is
satisfied. The TM and TE components of the output
light are separated by a polarization beam
splitter. The TM component corresponds to the
direct transmission of the pump light, while the
TE component contains the scattered sideband. (d)
Power dependence of the sideband on the input
microwave power $P_{\mathrm{MW}}$. Inset:
extracted sideband power as a function of the
input microwave power.}\label{fig:sideband}

\par\end{centering}

\end{figure}

To measure the interaction between optical photon
and magnon, the YIG sphere is biased at $H=2410$
Oe, corresponding to a magnon resonance frequency
of $\omega_\mathrm{m}/2\pi=6.75$ GHz. The
optomagnonic resonator is pumped by a TM
polarized laser beam with 1 mW power, and the
magnons are excited by an on-resonance microwave
signal. The laser wavelength is scanned to search
for the optical modes that satisfy the energy,
spin and angular momenta conservation conditions.
During the searching process, lock-in technique
is adopted to improve the converted light signal
to noise ratio. When the conservation conditions
are satisfied, the output light is sent to a high
resolution spectrometer for further analysis. It
is worth noting that the density of optical WGMs
is very large, as there are mode degeneracy in
the polar direction and high order modes in the
radial direction. As a result, the conservation
conditions can be satisfied accidentally, similar
to the Brillouin scattering in microsphere
optomechanical cavities \cite{Dong2015,Kim2015}.
A typical spectrum of converted photons as a
function of the sweeping pump laser wavelength is
shown in Fig.\,\ref{fig:sideband}(a), where the
passive transmission spectrum of the pump light
is also shown accordingly. The correspondence
between the resonances for pump light and the
peaks of magnon-photon conversion implies the
triple resonance enhancement in our optomagnonic
cavity. The dependence of the converted photons
on the microwave resonance
{[}Fig.\,\ref{fig:sideband}(b){]} also confirms
the participation of magnon in the inelastic
light scattering process.

The detailed spectrum for one selected
tripe-resonance condition is plotted in
Fig.\,\ref{fig:sideband}(c), where the
optomagnonic resonator is pumped by a TM light at
1534.599 nm. A polarization beam splitter is used
to separate the two polarizations. The TM
component of the output light shows a single peak
as it only contains the transmitted pump light.
On the contrary, the TE component shows two
peaks: a strong peak which corresponds to the
transmitted pump light that is not completely
filtered out, and a weak sideband which
corresponds to the generated photons. Therefore
we do have orthogonal polarizations for the
generated signal and the pump light, which agrees
well with our theory model. The linewidths of the
measured pump and sideband signal are not the
physical linewidth of the light but instead only
represent the finite resolution (67 MHz) of the
filter in the spectrometer. The centers of the
pump peak and sideband differ from each other by
$6.75$ GHz, matching the input magnon frequency.
The sideband appears only on one side of the pump
as a result of the conservation conditions, as
explained in above analysis. We measured the
converted light at various microwave input
powers, which clearly shows a linear power
dependence {[}Fig.\,\ref{fig:sideband}(d){]},
indicating a linear magnon to photon conversion.
The fitted raw power (system) conversion
efficiency is \textbf{$5\times10^{-8}$}.
Considering the imperfect resonance coupling and
in-line insertion losses for both the optical and
microwave circuits, the internal power conversion
efficiency is about $5\times10^{-3}$. The
conversion efficiency can be improved by using
YIG spheres of smaller size and smoother surface.
Further geometry optimization, such as using YIG
microdisk whose modal volume is orders of
magnitude smaller, in combination with Faraday
effect enhancement via doping, could lead to much
improved conversion efficiencies.

In conclusion, we have demonstrated an excellent
optomagnonic resonator that is made by a highly
polished YIG sphere. Utilizing an integrated
optical chip for high efficiency optical
coupling, high-$Q$ optical WGMs are observed in
addition to magnon resonances in the YIG sphere
after our careful surface treatment. When the
triple resonance condition and angular momentum
conservation condition are satisfied, the magnon
is converted to optical photon with internal
power efficiency of about $0.5\%$. This
efficiency can be further improved by doping or
geometry optimization. Our findings show that YIG
sphere is a promising platform for designing
complex hybrid systems, which holds great
potential to realize information inter-conversion
among magnon, microwave photon, and optical
photons.

The authors thank Liang Jiang and Michael
Flatt\'{e} for fruitful discussions, and funding
support from DARPA/MTO MESO program
(N66001-11-1-4114), a US Army Research Office
grant (W911NF-14-1-0563), an AFOSR MURI grant
(FA9550-15-1-0029), and the Packard Foundation.


\end{document}